\documentclass[amsmath,amssymb,aps,prd,nofootinbib,superscriptaddress,onecolumn,preprintnumbers]{revtex4-2}
\usepackage[utf8]{inputenc}
\usepackage{mathrsfs}
\usepackage{bm}
\usepackage{url}
\usepackage[normalem]{ulem}
\usepackage{mathtools}
\usepackage{array}
\newcolumntype{P}[1]{>{\centering\arraybackslash}p{#1}}
\newcolumntype{M}[1]{>{\centering\arraybackslash}m{#1}}
\usepackage[caption=false]{subfig}
\usepackage{dcolumn}
\usepackage{graphicx, epsfig}
\usepackage[dvipsnames]{xcolor}
\usepackage{mathrsfs}
\usepackage{bm}
\usepackage{yhmath}
\usepackage[caption=false]{subfig}
\usepackage[normalem]{ulem}
\usepackage{mathtools}
\usepackage{bigints}
\usepackage{float}
\usepackage[colorlinks = true, linkcolor = purple, urlcolor  = blue, citecolor = blue, anchorcolor = blue]{hyperref}
\usepackage{float}
\usepackage{multirow}
\usepackage{tikz,xcolor,hyperref}
\definecolor{darkgreen}{rgb}{0.0, 0.2, 0.13}
\definecolor{bostonuniversityred}{rgb}{0.8, 0.0, 0.0}
\definecolor{lime}{HTML}{A6CE39}
\DeclareRobustCommand{\orcidicon}{
	\begin{tikzpicture}
	\draw[lime, fill=lime] (0,0) 
	circle [radius=0.16] 
	node[white] {{\fontfamily{qag}\selectfont \tiny ID}};
	\draw[white, fill=white] (-0.0625,0.095) 
	circle [radius=0.007];
	\end{tikzpicture}
	\hspace{-2mm}
}
\foreach \x in {A, ..., Z}{\expandafter\xdef\csname orcid\x\endcsname{\noexpand\href{https://orcid.org/\csname orcidauthor\x\endcsname}
			{\noexpand\orcidicon}}
}
\newcommand{\be}{\begin{equation}}
\newcommand{\ee}{\end{equation}}
\newcommand{\ba}{\begin{eqnarray}}
\newcommand{\ea}{\end{eqnarray}}

\def\lp4{$\lambda \phi^4$}
\begin{document}
\preprint{YITP-25-27}

\title{Evolution of a kink-antikink ensemble in a quantum vacuum}
\author{Mainak Mukhopadhyay\hspace{-1mm}\orcidA{}} \email[Corresponding author: ]{mkm7190@psu.edu}
\affiliation{Department of Physics; Department of Astronomy \& Astrophysics; Center for Multimessenger Astrophysics, Institute for Gravitation and the Cosmos, The Pennsylvania State University, University Park, PA 16802, USA
}

\author{Kohta Murase\hspace{-1mm}\orcidB{}} \email{kmurase@psu.edu}
\affiliation{Department of Physics; Department of Astronomy \& Astrophysics; Center for Multimessenger Astrophysics, Institute for Gravitation and the Cosmos, The Pennsylvania State University, University Park, PA 16802, USA
}
\affiliation{Center for Gravitational Physics and Quantum Information, Yukawa Institute for Theoretical Physics, Kyoto University, Kyoto 606-8502, Japan}

\author{Atsushi Naruko\hspace{-1mm}\orcidC{}} \email{naruko@yukawa.kyoto-u.ac.jp}
\affiliation{Center for Gravitational Physics and Quantum Information, Yukawa Institute for Theoretical Physics, Kyoto University, Kyoto 606-8502, Japan}
\affiliation{Asia Pacific Center for Theoretical Physics, Pohang 37673, Korea}

\author{George Zahariade\hspace{-1mm}\orcidD{}} \email{george.zahariade@uj.edu.pl}
\affiliation{Instytut Fizyki Teoretycznej, Uniwersytet Jagiello\'nski, Lojasiewicza 11, 30-348 Krak\'ow, Poland}

\date{\today}

\begin{abstract}
We study the $1+1$ flat spacetime dynamics of a classical field configuration corresponding to an ensemble of sine-Gordon kinks and antikinks, semi-classically coupled to a quantum field. This coupling  breaks the integrability of the sine-Gordon model resulting in the background's decay into quantum radiation as kink-antikink pairs annihilate. We find evidence that, on average, the energy of the ensemble scales as $t^{-\alpha}$ with $\alpha<1$ and independent of the coupling strength or the mass of the quantum field. The generalization of this result to domain wall networks in higher spacetime dimensions could be relevant to particle production in the early universe. 
\end{abstract}
\maketitle
\section{Introduction}
\label{sec:intro}
Domain wall networks have been thoroughly studied in the context of primordial cosmology, since such configurations generically form when a discrete symmetry is spontaneously broken (leading to a non-trivial vacuum manifold with potentially multiple degenerate vacua) and their presence can have dramatic consequences on the history of the early universe~\cite{Zeldovich:1974uw,Vilenkin:2000jqa,Vachaspati:2006zz}. Numerous analytical~\cite{Hindmarsh:1996xv,Hindmarsh:2002bq} and numerical~\cite{Press:1989yh,Larsson:1996sp,Garagounis:2002kt,Oliveira:2004he,Hiramatsu:2010yz,Leite:2011sc,Kawasaki:2011vv,Hiramatsu:2013qaa,Correia:2014kqa,Correia:2018tty,ZambujalFerreira:2021cte,Kitajima:2023cek,Chang:2023rll,Ferreira:2024eru} studies have pointed out that domain wall networks reach a so-called {\it scaling regime} where their decay is governed by a power law i.e. where the energy stored in the domain wall network scales as $t^{-\alpha}$ with $\alpha>0$.
Indeed, assuming that the domain wall network is seeded via random but unbiased initial conditions (where every point in space has equal probability to pick either vacuum), the correlation length dynamically increases until reaching the maximum value allowed by causality i.e. $H^{-1} \sim t$ in Friedmann-Lema\^itre-Robertson-Walker (FLRW) spacetime. This implies a domain wall area density $\sim t^{-1}$ where there is on average one domain wall per Hubble patch. The existence of this \emph{linear} scaling regime was established in several works for Minkowski space, as well as for radiation-dominated and matter-dominated FLRW models~\cite{Press:1989yh,Hindmarsh:1996xv,Larsson:1996sp,Hindmarsh:2002bq,Garagounis:2002kt,Oliveira:2004he,Hiramatsu:2010yz}. These conclusions were solidified with more precise numerical computations where previous claims of deviation from the linear scaling regime were attributed to limited dynamical range (or size of the physical box)~\cite{Leite:2011sc,Kawasaki:2011vv,Hiramatsu:2013qaa}. It was also pointed out that the presence of a bias in the initial conditions leads to an additional exponential suppression~\cite{Hindmarsh:1996xv,Hiramatsu:2010yz} of the domain wall area density. A similar, albeit not identical, suppression factor was also shown to arise in the {\it pressure biased} case (with non-degenerate distinct vacua after symmetry breaking)~\cite{Correia:2014kqa,Correia:2018tty,Kitajima:2023cek}.

Since domain walls are usually treated as classical field configurations, most of these studies have been conducted in a fully classical setting, either by assuming effective Nambu-Goto dynamics, or by describing the networks in a classical field theoretic framework. It should be mentioned that there exist regimes (the so-called {\it spinodal instability}, or {\it domain wall precursor}, regime) where a full quantum treatment is possible~\cite{CALZETTA198932,Boyanovsky:1993fy,Ibaceta:1998yy,Boyanovsky:1999wd,Mukhopadhyay:2020xmy,Mukhopadhyay:2020gwc,Pujolas:2022qvs,Mukhopadhyay:2023zmc,Mukhopadhyay:2024wii} but it is unclear to what extent this is relevant to realistic cosmological scenarios.

In this work, we pursue a hybrid approach in which the domain wall network is treated as a classical field configuration $\phi(t,x)$ whose main decay channel is via particle-antiparticle pair production in a second field $\psi(t,x)$ treated quantum mechanically. Similar studies investigating the production of mesons in kink-antikink interactions were performed in ~\cite{Evslin:2022fzf,Evslin:2023egm,Evslin:2023oub}. The coupling between the two fields is understood to be semi-classical and we further assume that the network's annihilation due to its self-interactions can be neglected. The aim is to investigate whether a scaling regime exists in such a semi-classical setting and to find the corresponding power law~$\alpha$. 

For reasons of computational complexity we limit ourselves to a $1+1$ dimensional setting where the domain wall network reduces to an ensemble of kinks and antikinks. We further assume a sine-Gordon potential $V(\phi)\propto 1-\cos\phi$ for the broken symmetry field $\phi$, along with a $V(\phi)\psi^2$ coupling to the quantum field $\psi$. The integrability properties of the sine-Gordon model in $1+1$ dimensions ensure that the decay of the field configuration primarily occurs via quantum radiation of $\psi$ quanta. The model and its evolution equations will be briefly described in Sec.~\ref{sec:formalism}. The setup is essentially the same as in Ref.~\cite{Mukhopadhyay:2021wmu} where the scattering of one kink-antikink pair in the presence of quantum radiation was studied in detail and the outcome of the collision (annihilation or elastic scattering) was found to depend finely on the relative velocity. Here we will rather be interested in the time evolution of global properties such as the total energy of a kink-antikink ensemble, its kink number density or, the total radiated energy. For a sufficiently large number of kink-antikink pairs we expect the late time behavior of such quantities to lose memory of the specific initial conditions and maybe even exhibit universal properties. In Sec.~\ref{sec:numerical_sol} we put this hypothesis to the test by simulating the dynamics of several ensembles of kink-antikink pairs with randomly generated relative positions and velocities. Averaging over these distinct simulations yields a late time scaling law for the total energy in the kink-antikink background that seems to slightly disagree with the classical one, $\alpha=1$, that is usually considered in the literature (see e.g.~\cite{Hindmarsh:1996xv}). We conclude in Sec.~\ref{sec:disc} with a discussion of the limitations and possible improvements of our results as well as their relevance to cosmology.

\section{Setup}
\label{sec:formalism}
The action describing the full dynamics of our $1+1$ dimensional system is\footnote{We work in units where $\hbar=c=1$.}
\be
S=\int dt dx \left[\frac{1}{2}\dot{\phi}^2-\frac{1}{2}\phi'^2 - m^2(1-\cos
\phi)+\frac{1}{2}\dot{\psi}^2-\frac{1}{2}\psi'^2 -\frac{1}{2}\mu^2\psi^2-\frac{\lambda}{2} (1-\cos\phi) \psi^2\right],
\label{eq:fullmodel}
\ee
where we use the standard dot and prime notations for temporal and spatial derivatives respectively. The two fields $\phi$ and $\psi$ have potentially different masses $m$, $\mu$ and their coupling is denoted by $\lambda$.

We will begin by setting up the field profile corresponding to the initial configuration of kink-antikink pairs, $\phi_0(x)$. One sine-Gordon kink (or antikink) has the well-known form
\be
\phi_{\pm}(t,x)=\pm 4 \arctan\left[e^{m\gamma(x-x_0-vt)}\right],
\ee
where $x_0$ represents its position at $t=0$ and $\gamma$ is the Lorentz contraction factor associated with its velocity $v$. The overall sign distinguishes between a kink and an antikink. We can use this class of exact solutions of the $\phi$ field equations for $\lambda=0$ to build the desired ensemble of $n$ kink-antikink pairs by simply pasting together (at time $t=0$ say) $2n$ alternating kink and antikink profiles centered at positions $x_i$ (separated by distances larger than the typical kink width, $|x_{i+1}-x_i|\gg1/m$), and with randomly generated velocities $v_i$. This defines the initial field profile
\be
\phi_0(x) = 4 \sum_{i=1}^{2n} (-1)^{i + 1} {\rm arctan} \left[ e^{m \gamma_i (x - x_i)} \right],
\label{eq:phiIC}
\ee
along with the associated time derivative profile
\be
\dot{\phi}_0(x)=2m\sum_{i=1}^{2n} (-1)^{i} \gamma_i v_i\, {\rm sech} \left[ m \gamma_i (x - x_i) \right].
\label{eq:phidotIC}
\ee
We choose such an alternating kink-antikink configuration (instead of more general ones with multiple adjacent kinks or antikinks) to ensure our results also apply to situations with only two degenerate vacua such as the double well potential. This has the added advantage of making it easy to satisfy periodic boundary conditions. 

In the absence of interactions with the quantum radiation field $\psi$ ($\lambda=0$) the subsequent dynamics of the kink-antikink ensemble is given by the $1+1$-dimensional sine-Gordon equation
\be
\ddot{\phi}-\phi^{\prime\prime}+m^2\sin{\phi}=0,
\label{eq:sG}
\ee
whose integrability properties ensure that the system's evolution reproduces a {\it stable} multiple-kink-antikink solution.\footnote{This holds as long as the initial separations are large enough. For arbitrary separations, one should use the corresponding analytical solution of the sine-Gordon equation, obtained by using the B\"acklund transform~\cite{1990nwsc.book.....I}.} This system thus verifies the assumption of negligible self-annihilation: the kinks and antikinks simply pass through each other without decaying, either forming bound states or scattering to infinity. In fact the entsemble does not annihilate at all.

The situation changes radically when the interaction with the quantum radiation field is turned on ($\lambda\neq 0$). Now the integrability of the sine-Gordon equation is broken and the kink-antikink ensemble can decay into $\psi$ quanta. Intuitively, non-negligible particle production in this channel will occur when the kink-antikink ensemble, acting as a dynamical background for the quantum field $\psi$, ceases to be adiabatic, i.e., whenever a kink-antikink collision occurs. The energy of the emitted $\psi$ quanta can only come from the background which then decays. Such a situation is a simple example of backreaction of quantum radiation on a classical background.

To study how this happens we shall work within a semi-classical approximation, where the backreaction of the produced $\psi$ quanta on the background dynamics is given by a modification of~Eq.~\eqref{eq:sG}:
\be
\ddot{\phi}-\phi^{\prime\prime}+\left[m^2+\frac{\lambda}{2}\langle\psi^2\rangle\right]\sin{\phi}=0.
\label{eq:sGbar}
\ee
This can be obtained by varying action~\eqref{eq:fullmodel} with respect to $\phi$ and replacing any operator depending on the quantum field $\psi$ in the resulting field equation by a quantum expectation value. 

One then still needs to solve for the dynamics of $\langle\psi^2\rangle$. This is most readily accomplished in the Heisenberg picture where the quantum operator $\psi$ obeys the (linear) field equation
\be
\ddot{\psi}-\psi^{\prime\prime}+\left[\mu^2+\lambda(1-\cos{\phi})\right]\psi=0,
\label{eq:qrad}
\ee
where $\phi(t,x)$ is treated as a given classical background. This equation is also straightforwardly obtained from varying action given by Eq.~\eqref{eq:fullmodel} with respect to $\psi$. Specifying the quantum state of the field $\psi$ to be the zeroth order adiabatic vacuum, $|0\rangle$, given by the background field configuration at $t=0$, $\phi(0,x)=\phi_0(x)$, one can solve for the dynamics of $\langle\psi(t,x)^2\rangle\equiv\langle 0|\psi(t,x)^2|0\rangle$ using any of the tried and tested methods developed for computing particle production in time dependent backgrounds. Then, plugging this solution back in Eq.~\eqref{eq:sGbar}, we obtain the desired backreacted dynamics for the classical kink-antikink ensemble background $\phi(t,x)$.

We shall make these statements more precise by using the methods introduced and described in detail in Refs.~\cite{Vachaspati:2018llo,Vachaspati:2018hcu,Mukhopadhyay:2019hnb,Mukhopadhyay:2021wmu,Mukhopadhyay:2023zmc,Mukhopadhyay:2024wii}. The first step is to discretize the problem on a regular $N$ point lattice of size $L$ with periodic boundary conditions. The lattice spacing is then given by $a=L/N$ and, using a three point stencil to discretize the spatial second derivative operator, Eqs.~\eqref{eq:sGbar} and~\eqref{eq:qrad} become 
\ba
&&\ddot{\phi}_i-\frac{1}{a^2}\left(\phi_{i+1}-2\phi_i+\phi_{i-1}\right)+\left[m^2+\frac{\lambda}{2}\langle\psi_i^2\rangle\right]\sin{\phi_i}=0,
\label{eq:discsGbar}\\
&&\ddot{\psi}_i-\frac{1}{a^2}\left(\psi_{i+1}-2\psi_i+\psi_{i-1}\right)+\left[\mu^2+\lambda(1-\cos{\phi_i})\right]\psi_i=0,
\label{eq:discqrad}
\ea
where the $\phi_i$ and $\psi_i$ are the (time-dependent) values of the classical field $\phi$ and quantum operator $\psi$ at the $i$-th lattice point $-L/2+ia$.
Assembling the $\psi_i$ field values into a length $N$ column vector $\bm{\psi}$ we can rewrite Eq.~\eqref{eq:discqrad} in the more compact way
\be
\ddot{\bm{\psi}}+\bm{\Omega}_\lambda^2.\bm{\psi} = 0,
\ee
where the $N\times N$ matrix $\bm{\Omega}_\lambda^2$ is defined by
\be
[\bm{\Omega}_\lambda^2]_{ij} = 
\begin{cases}
+{2}/{a^2}+\mu^2+\lambda\left(1-\cos{\phi}_i\right)&\text{for}\ i=j\\
-{1}/{a^2}&\text{for}\ i=j\pm1\ (\text{mod}\ N)\\
0&\text{otherwise}\,.
\end{cases}
\label{Omega2}
\ee
Following Refs.~\cite{Vachaspati:2018llo,Vachaspati:2018hcu,Mukhopadhyay:2019hnb,Mukhopadhyay:2021wmu,Mukhopadhyay:2023zmc,Mukhopadhyay:2024wii} we know that we can solve for the quantum dynamics of the field operator $\psi$ by solving for a complex $N\times N$ matrix, $\bm{Z}$, verifying
\be
\ddot{\bm{Z}}+\bm{\Omega}_\lambda^2.\bm{Z}=0,
\label{eq:CQCeqs}
\ee
with initial conditions 
\be
\bm{Z}(t=0)=-\frac{i}{\sqrt{2a}}\bm{\Omega}_\lambda(t=0)^{-1/2}\quad\text{and}\quad \dot{\bm{Z}}(t=0)=\frac{1}{\sqrt{2a}}\bm{\Omega}_\lambda(t=0)^{1/2},
\label{eq:CQCics}
\ee
for our specific choice of vacuum $|0\rangle$. Indeed all two-point functions can be computed from $\bm{Z}$ as
\be
\langle\bm{\psi}.\bm{\psi}^T\rangle= \bm{Z}.\bm{Z}^\dag,\quad\langle\dot{\bm{\psi}}.\bm{\psi}^T\rangle= \dot{\bm{Z}}.\bm{Z}^\dag,\quad\langle{\bm{\psi}}.\dot{\bm{\psi}}^T\rangle= {\bm{Z}}.\dot{\bm{Z}}^\dag,\quad\langle\dot{\bm{\psi}}.\dot{\bm{\psi}}^T\rangle= \dot{\bm{Z}}.\dot{\bm{Z}}^\dag.
\label{eq:2point}
\ee
The middle two equations can be shown~\cite{Vachaspati:2018hcu} to be consistent with the canonical commutation relations verified by $\psi$ and $\dot{\psi}$. Since $\psi$ has no self-interactions, knowledge of the two-point functions determines the full quantum dynamics.

There remains however a subtlety. While the above analysis is exact in the discrete case, the two-point function of interest, $\langle \psi_i^2\rangle$, involves field values at the same point and this is known to suffer from divergences in the continuum limit ($N\to\infty$, $a\to 0$ with $L$ kept fixed). We therefore work with {\it renormalized} two-point functions obtained by subtracting from Ref.~\eqref{eq:2point} the trivial vacuum contribution i.e. the value of the corresponding two-point functions obtained from the solution, 
\be
\bar{\bm{Z}}=-\frac{i}{\sqrt{2a}}\bm{\Omega}_0^{-1/2}\exp{i\bm{\Omega}_0t}, 
\ee
of Eq.~\eqref{eq:CQCeqs} for $\lambda=0$ (or equivalently $\phi_i=0$). The renormalized two point functions $\psi\psi$ and $\dot{\psi}\dot{\psi}$ will for instance read
\ba
\langle\bm{\psi}.\bm{\psi}^T\rangle_R&=& \bm{Z}.\bm{Z}^\dag-\bar{\bm{Z}}.\bar{\bm{Z}}^\dag=\bm{Z}.\bm{Z}^\dag-\frac{1}{2a}\bm{\Omega}_0^{-1},\\
\langle\dot{\bm{\psi}}.\dot{\bm{\psi}}^T\rangle_R&=& \dot{\bm{Z}}.\dot{\bm{Z}}^\dag-\dot{\bar{\bm{Z}}}.\dot{\bar{\bm{Z}}}^\dag=\dot{\bm{Z}}.\dot{\bm{Z}}^\dag-\frac{1}{2a}\bm{\Omega}_0,
\ea
with analogous definitions for the $\psi\dot{\psi}$ and $\dot{\psi}\psi$ cases (that we won't be needing in the following). One can check that this definition of the two-point functions is stable against changes in the coarseness of the lattice provided $a$ is small enough~\cite{Mukhopadhyay:2021wmu}. In particular this procedure provides a $\bm{Z}$-dependent expression for the two-point function $\langle\psi_i^2\rangle$ appearing in Eq.~\eqref{eq:discsGbar}, that is UV stable, namely
\be
\langle\psi_i^2\rangle_R=\sum_{j=1}^N\left(|Z_{ij}|^2-|\bar{Z}_{ij}|^2\right),
\ee
where we have introduced the components of $Z_{ij}$, $\bar{Z}_{ij}$ of the complex matrices $\bm{Z}$ and $\bar{\bm{Z}}$.\footnote{This definition tacitly implies that the parameter $m$ appearing in Eq.~\eqref{eq:sGbar} is the physical mass of $\phi$. The subtraction is the result of the $\phi$ field mass renormalization.}

In summary, we can find the semiclassically backreacted dynamics of the discretized system by numerically solving the $N^2$ ordinary differential equations for the quantum field $\psi$ already given in Eq.~\eqref{eq:CQCeqs} with initial conditions~\eqref{eq:CQCics}, along with the $N$ ordinary differential equations for the classical background field $\phi$, 
\be
\ddot{\phi}_i-\frac{1}{a^2}\left(\phi_{i+1}-2\phi_i+\phi_{i-1}\right)+\left[m^2+\frac{\lambda}{2}\sum_{j=1}^N\left(|Z_{ij}|^2-|\bar{Z}_{ij}|^2\right)\right]\sin{\phi_i}=0,
\label{eq:discsGbarbis}
\ee
with initial conditions given by~\eqref{eq:phiIC}
and~\eqref{eq:phidotIC}. In the next section we will use this method to investigate numerically how the quantum decay of a generic kink-antikink ensemble occurs.

\section{Numerical results}
\label{sec:numerical_sol}

The main aim of this work is to investigate the late time dynamics of a large kink-antikink ensemble prepared at time $t=0$ in a random fashion, i.e., with random relative positions and velocities. More precisely the positions $x_i$ and velocities $v_i$ are  sampled from uniform distributions over the intervals $[-L/2,L/2]$, $[-v_{max},v_{max}]$ respectively. The maximum velocity $v_{max}$ is chosen so as to not introduce too much non-adiabaticity at the initial time, and in practice we shall choose $v_{max}=0.4$. Using the discretized formalism introduced in the previous section, we would therefore have to choose a lattice of size $L$ large enough to accommodate a large number $n$ of well-separated kink-antikink pairs at $t=0$ ($L\gg 2n/m$). We would also need to choose a large enough number $N$ of lattice points to be able to resolve the individual kinks and antikinks ($a\ll 1/m$ or $N\gg mL$). We then would build an initial kink-antikink ensemble configuration $\phi_0(x)$ as in~\eqref{eq:phiIC}, and finally evolve it by solving the coupled equations~\eqref{eq:CQCeqs} and~\eqref{eq:discsGbarbis} with initial conditions given by~\eqref{eq:CQCics} and~\eqref{eq:phiIC}, \eqref{eq:phidotIC} respectively.

However we immediately get faced with the computational limitations of our setup since the algorithmic complexity scales with $N^2$ i.e. proportionally to the number of components of the matrix $\bm{Z}$. This places an upper bound on the number of kinks and antikinks we can include in our ensemble. In practice this means we will be limited to $n\lesssim\mathcal{O}(10)$ and we can expect our results to depend strongly on the particular initial configuration $\phi_0$ of the kink-antikink ensemble. Despite this caveat, we still hope to obtain universal results by repeating the simulation for different choices of $\phi_0$ and taking an average of the corresponding numerical solutions for $\phi(t,x)$.

\subsection{A typical example simulation}
\label{subsec:one_sol}

\begin{figure}
\begin{center}
\includegraphics[width=0.7\textwidth,angle=0]{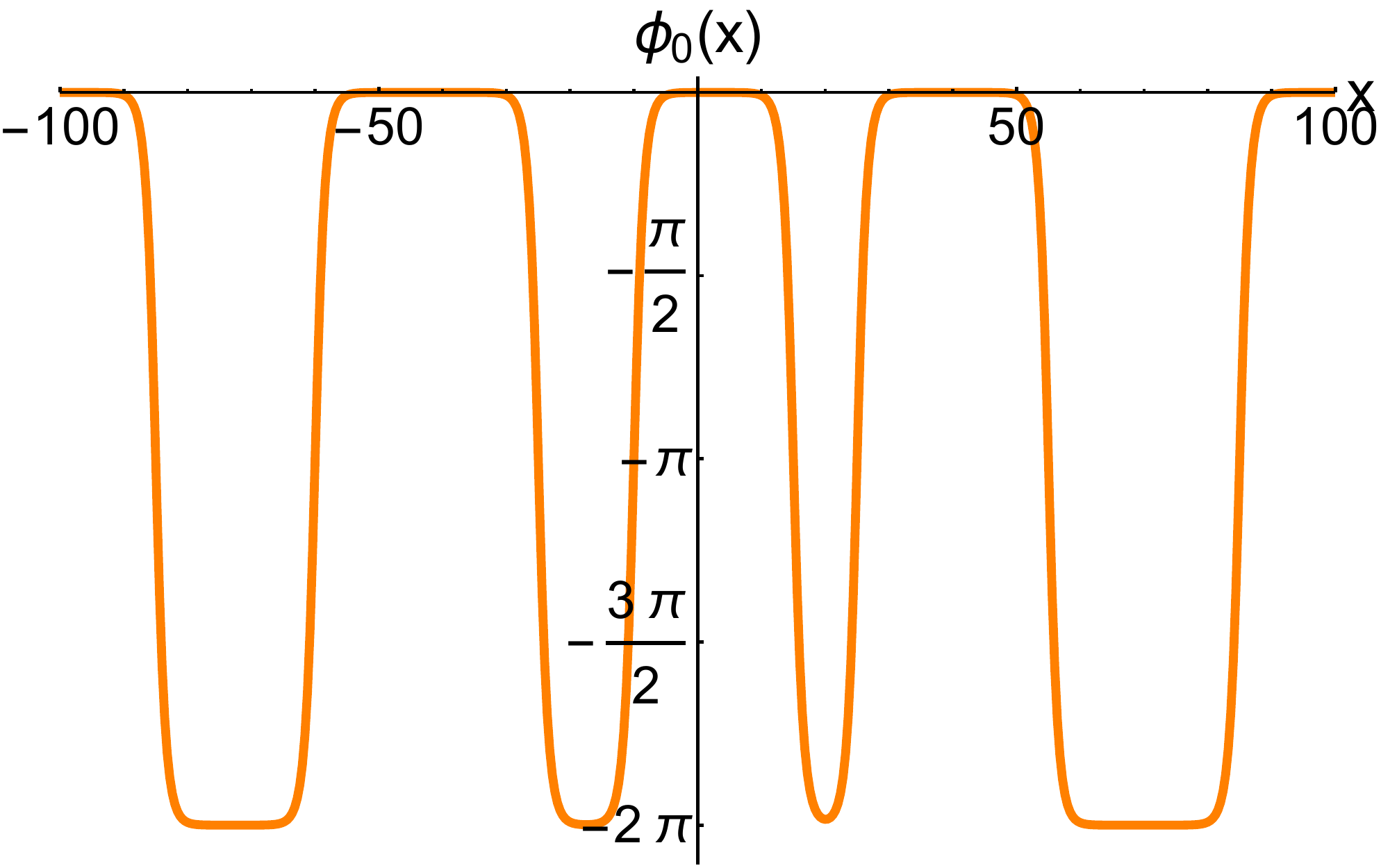}
\caption{\label{fig:phibkg} One possible initial configuration for the kink-antikink ensemble background, $\phi_{0}(x)=\phi(t=0,x)$. The units are such that $m=1$.
}
\end{center}
\end{figure}

\begin{figure}
\begin{center}
\includegraphics[width=0.75\textwidth,angle=0]{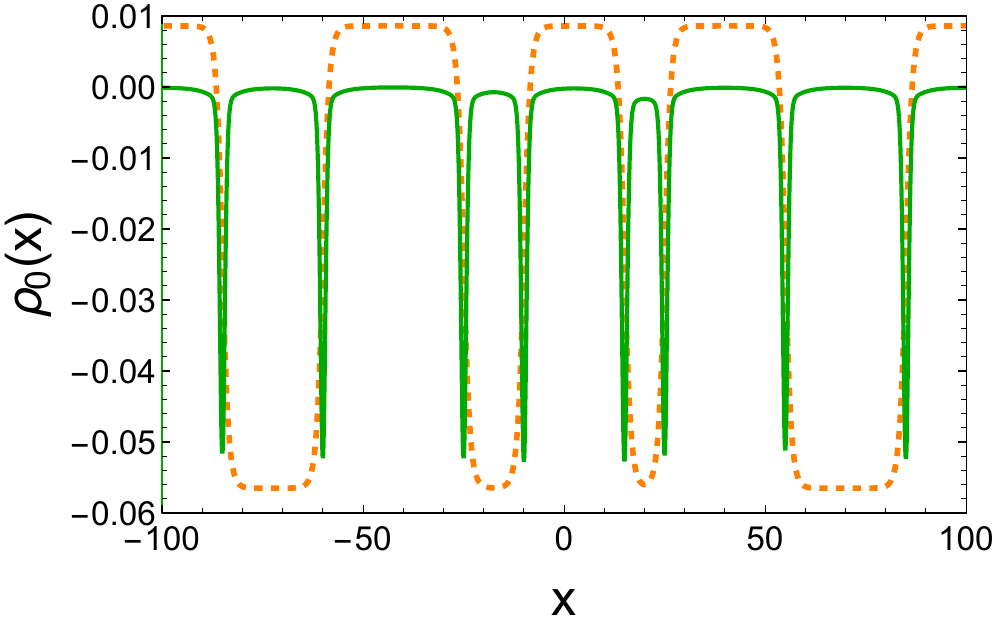}
\caption{\label{fig:rho} Initial profile of the renormalized average energy density in the quantum field $\psi$, $\rho_0(x)=\rho(t=0,x)$, of which $\rho_i$ is the discretized version. The classical background $\phi_0$ is shown in dashed lines for reference. The kinks and antikinks are dressed by clouds of virtual $\psi$ particles that appear to have negative energy density. Here $\lambda=0.3$, $\mu=0.1$, and the units are such that $m=1$.
}
\end{center}
\end{figure}

One simulation will involve setting up a particular initial kink-antikink ensemble profile $\phi_0(x)$ as illustrated, e.g., in Fig.~\ref{fig:phibkg}. This will constitute the initial conditions for the classical field $\phi(t,x)$ (and its discretized version $\phi_i$). 
The corresponding initial conditions for the $\bm{Z}$ variable are harder to visualize but they fix the initial value of the discretized (renormalized) average energy density in the quantum field $\psi$,
\ba
\rho_i&=& \sum_{j=1}^N\left( \frac{1}{2}|\dot{Z}_{ij}|^2+\frac{1}{4a^2}\left[|Z_{i+1,j}-Z_{ij}|^2+|Z_{ij}-Z_{i-1,j}|^2\right]+\frac{1}{2}\left\{\mu^2+\lambda\left[1-\cos{\phi}_i\right]\right\}|Z_{ij}|^2\right)\nonumber\\
&&-\ (\text{same quantity with $Z\to \bar{Z}$})\,.
\ea
Notice that, for numerical accuracy reasons, the discretized gradient term was computed as an average of forward and backward differences. We plot this initial energy density for the choice of parameters $\lambda=0.3m^2$, $\mu=0.1m$ in Fig.~\ref{fig:rho}.

With this initial setup, we are ready to numerically evolve the differential equations~\eqref{eq:CQCeqs} and ~\eqref{eq:discsGbarbis}. We work in units where $m=1$ and choose a lattice of size $L = 200$ with $N = 500$ points. We use a position \emph{Verlet} method~\cite{Mukhopadhyay:2023zmc} with a time step $dt = a/5$, which leads to total energy conservation with $\sim 0.1$\% accuracy for the duration of our time evolution. As discussed at length in Ref.~\cite{Mukhopadhyay:2021wmu}, the implementation of the renormalization scheme makes our numerical results largely UV- and IR-independent, i.e., insensitive to the choices of $N$ and $L$.

\begin{figure}
\begin{center}
\includegraphics[width=0.75\textwidth,angle=0]{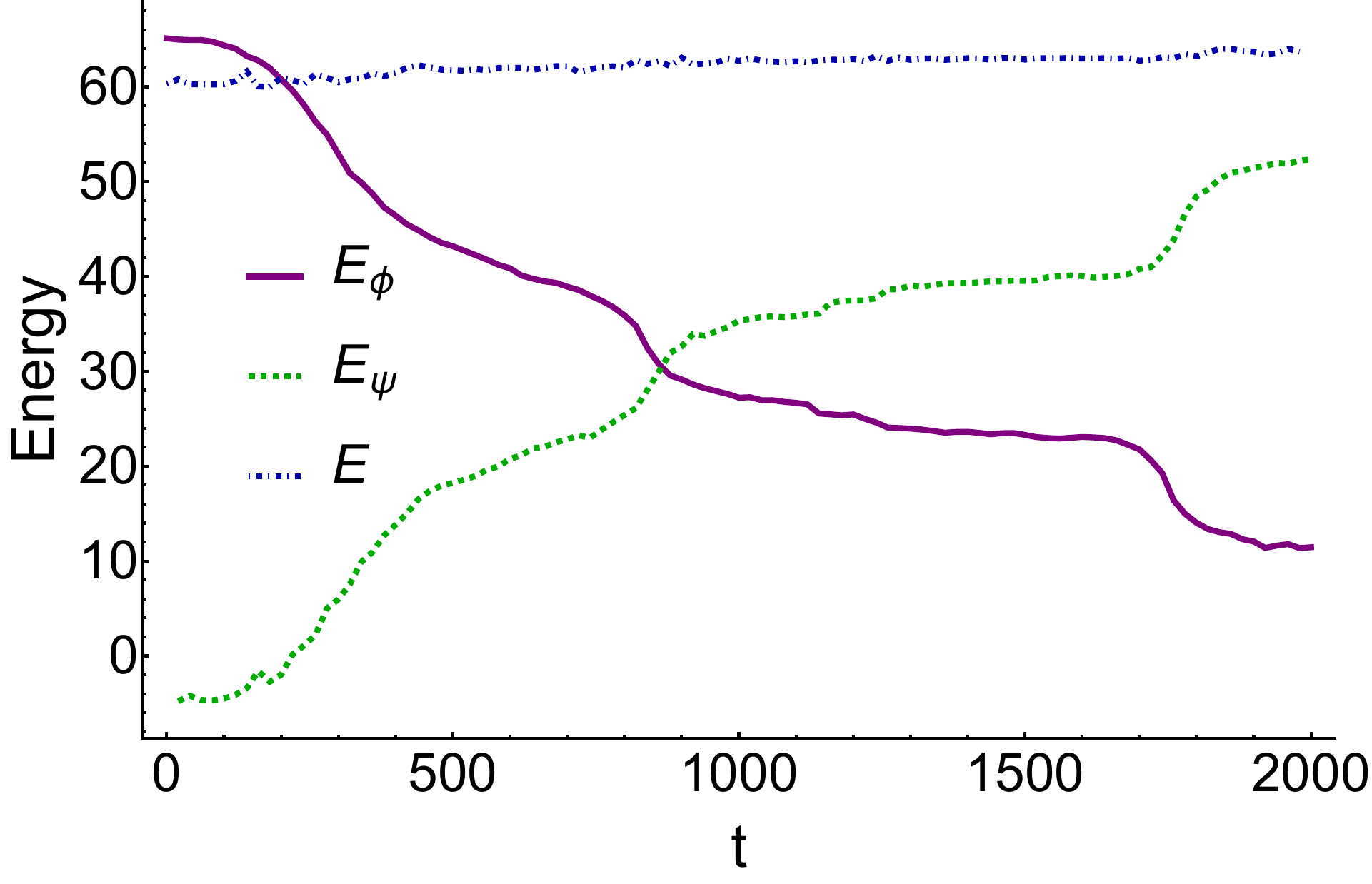}
\caption{\label{fig:energy_1sim} Time evolution of the energy in the background (solid line), the renormalized energy in the quantum field (dashed line), as well their sum (dot-dashed dark blue line). Here $\lambda=0.3$, $\mu=0.1$, and the units are such that $m=1$.
}
\end{center}
\end{figure}

In this work, we are in particular interested in the time variation of the energy in the classical kink-antikink ensemble,
\be
E_\phi = a\sum_{i=1}^N \left[\frac{1}{2}\dot{\phi}_i^2+\frac{1}{2a^2}\left(\phi_{i+1}-\phi_{i}\right)^2+ m^2(1-\cos
\phi_i)\right],\,.
\ee
Our result is shown in Fig.~\ref{fig:energy_1sim} (for the previous choice of parameters, $\lambda=0.3$, $\mu=0.1$) along with the (renormalized) energy in the quantum field $\psi$,
\be
E_\psi = \frac{a}{2}{\rm Tr}\left[\dot{\bm{Z}}.\dot{\bm{Z}}^\dag-\dot{\bar{\bm{Z}}}.\dot{\bar{\bm{Z}}}^\dag+\bm{\Omega}_\lambda^2.\left(\bm{Z}.\bm{Z}^\dag-\bar{\bm{Z}}.\bar{\bm{Z}}^\dag\right)\right].
\ee
These expressions are somewhat ambiguous because in the presence of interactions there is no good way of defining the energy of a subsystem. However it is clear that both $E_\phi$ and $E_\psi$ reduce to the energy in $\phi$ and to the quantum average of the energy in $\psi$ (up to an additive constant) respectively, when $\lambda=0$. 
Moreover, by using the field equations, we can readily check that the total energy, $ E = E_\phi+E_\psi$ is a conserved quantity (even when $\lambda\neq 0$). We therefore expect these definitions to correspond to the intuitive notions of ``energy of the classical background'' and ``energy in the quantum radiation'' at least for weak coupling and as long as the background stays classical. It is worth to pause for a moment here to discuss what exactly it means to ``stay classical.'' In fact our whole analysis is predicated upon the field $\phi$ being classical but this cannot possibly be true for all time. Much has been said about this topic in the literature~\cite{Dvali:2017eba,Dvali:2017ruz,Dvali:2022vzz,Michel:2023ydf,vanDissel:2024peo} but in this work we will adopt a naive criterion for classicality of $\phi$: we will assume that as long as the energy in the $\psi$ excitations (i.e., the renormalized energy) is less than the energy in $\phi$, $E_\psi\lesssim E_\phi$, the background can be treated classically and the semi-classical approximation holds. 

Since our numerical setup trades the full field theory for a system with finite degree of freedom system, we also would expect the radiated energy in $\psi$ to eventually return to the classical background. However for large enough $N$ (and in particular for the values considered in this work) this is expected, and can be verified, to happen for a Poincar\'e recurrence time far beyond the time duration of our evolution.

In Fig.~\ref{fig:energy_1sim} we can see how the classical kink-antikink ensemble background decays into quantum particles $\psi$. After a short initial phase during which $E_\phi$ is approximately constant (before kink-antikink collisions start occurring), and during which $E_\psi$ is negative (a sign that no {\it real} $\psi$ quanta are present), the decay of the classical background occurs approximately step-wise: for each kink-antikink collision, a {\it real} $\psi$ wavepacket is produced and emitted~\cite{Mukhopadhyay:2021wmu}, which leads to a sudden decrease in $E_\phi$ (and a corresponding increase in $E_\psi$). This discrete decay is however smoothed out to some extent by virtue of multiple collisions happening in close succession. Had we been able to run the simulation with a larger kink-antikink ensemble, the discrete nature of the decay would presumably be completely lost. We notice that, as expected, the total energy of the system is conserved to a good approximation, which increases our confidence in the stability of our numerical calculations (see Appendix~\ref{appsec:numerical_tech} for other consistency tests of our numerical methods).

\subsection{Average over many simulations}
\label{subsec:average_sol}

As mentioned before, our main limitation resides in the fact that we cannot evolve a large enough kink-antikink ensemble. Indeed this does not allow us to identify a potential late-time scaling regime where the number density of kinks and antikinks would follow a smooth power law. To try to get around this, we evolve multiple {\it small} kink-antikink ensembles such as the one described in Sec.~\ref{subsec:one_sol} for different, randomized, choices of the initial kink and antikink relative positions and velocities. The idea is to then plot the average energy in the classical kink-antikink ensemble background, $E_\phi$, over all these realizations. This should provide us with a {\it smoothed out} measure of the average decay dynamics of the kink-antikink ensemble.
\begin{figure}
\begin{center}
\includegraphics[width=0.75\textwidth,angle=0]{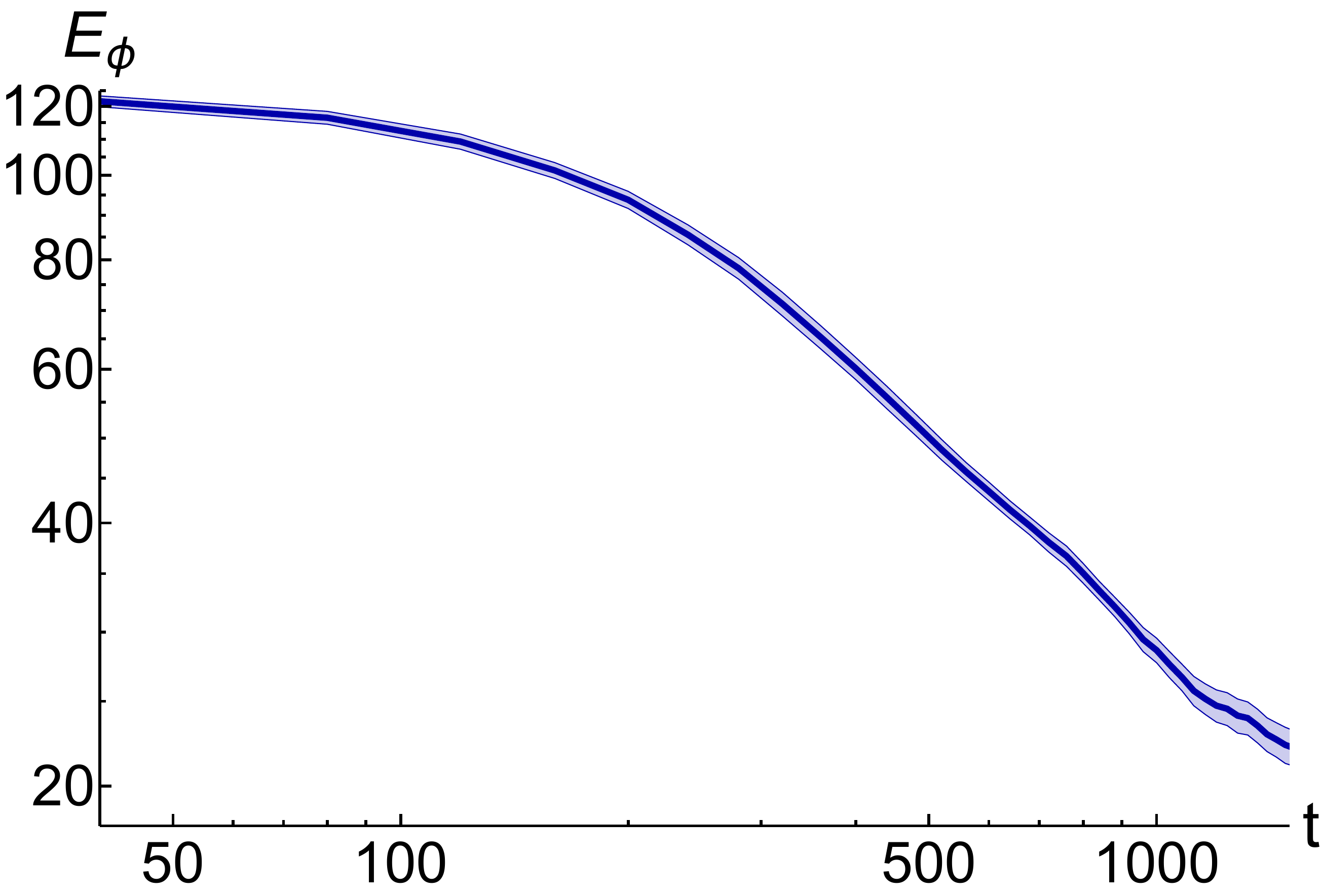}
\caption{\label{fig:main_res} Time-evolution of the energy in the background $E_\phi$ for an ensemble of $n = 8$ kink-antikink pairs averaged over a sample of $N_S=20$ different realizations, corresponding to 20 different, randomly chosen, initial conditions. 
Here $\lambda=0.3$, $\mu=0.1$, and the units are such that $m=1$. The shaded band shows the statistical uncertainty given by $\pm (\sigma/\sqrt{N_s})$, where $\sigma$ is the standard deviation of the $N_s$ samples.
}
\end{center}
\end{figure}

We choose the same parameters as above ($\lambda=0.3$, $\mu=0.1$) and consider a sample of $N_S=20$ different initial configurations of a kink-antikink ensemble with $n=8$ kink-antikink pairs. Each of these field configurations is generated by choosing the positions $x_i$ of the kinks and antikinks (see Eq.~\eqref{eq:phiIC}) to be uniformly sampled from the set $\{-L/2+ia\,|\, 1\leq i\leq N\}$ with the additional constraint that they shouldn't lie closer than $4/m$ to one another (or to the edges of the lattice, to automatically ensure compatibility with periodic boundary conditions).\footnote{Because of the spacing constraint, the $x_i$ values are sampled from a  distribution which is not strictly uniform.} This constraint is meant to ensure that $\phi_0(x)$ obeys periodic boundary conditions and that the kink-antikink ensemble is initially non-interacting. Similarly, up to a randomly chosen $\pm$ sign, the velocities $v_i$ (see Eq.~\eqref{eq:phidotIC}) are chosen to be uniformly sampled from the real interval $[0.05,0.4]$. The lower bound is meant to prevent our ensemble from taking too long to decay while the upper bound is meant to ensure that our choice of state for the quantum field $\psi$ (see Eq.~\eqref{eq:CQCics}) does not introduce large non-adiabaticity at $t=0$ \cite{Mukhopadhyay:2021wmu}. The time evolution of $E_\phi$, averaged over these realizations, is shown on a Log-Log scale in Fig.~\ref{fig:main_res}. It is clear that for times $t\gtrsim 300m^{-1}$, in an average sense, the energy in the kink-antikink ensemble decays following a power law, 
\be
E_\phi\propto t^{-\alpha},
\ee
We can indeed do a fit through the asymptotic tail of Fig.~\ref{fig:main_res} to determine the exponent $\alpha$. Since we can only trust our {\it classical} background approximation for the field $\phi$ as long as $E_\phi\gtrsim E_\psi$, the time range, $[t_{\rm start}, t_{\rm end}]$, of our fit needs to be such that $t_{\rm end}$ verifies $E_{\phi}(t_{\rm end})\gtrsim E_{\psi}(t_{\rm end})\sim E_{\phi}(0)/2$. Due to the limited size of our statistical sample ($N_s=20$), the computed value of the exponent $\alpha$ has an associated uncertainty which can be estimated using the following equation
\be
\label{eq:error}
\delta \alpha = \frac{ \delta \log{ E_{\rm start}} + \delta\log{E_{\rm end}}}{\log{t_{\rm end}} - \log{t_{\rm start}}}=\frac{ \delta E_{\rm start}/E_{\rm start} + \delta E_{\rm end}/E_{\rm end}}{\log{t_{\rm end}} - \log{t_{\rm start}}}\,,
\ee
where $E_{\rm start}$ and $E_{\rm end}$ are the values of $E_\phi$ at times $t_{\rm start}$ and $t_{\rm end}$ respectively. The symbol $\delta$ denotes the absolute uncertainty. Moreover $E_\phi$ is obtained by averaging over the $N_s$ realizations, so we can estimate $\delta E_{\rm start}$ and $\delta E_{\rm end}$ by the corresponding sample standard deviations divided by $\sqrt{N_S}$. We find
\be
\alpha=0.63 \pm 0.08\,,
\ee
for this specific parameter set. Notice that since $E_\phi+E_\psi$ is constant the scaling parameter $\alpha$ also controls the average rate of quantum particle production in the field $\psi$ at late times. We also show a statistical uncertainty band for the curve in Fig.~\ref{fig:main_res}. This is estimated by considering a deviation of $\pm \sigma/\sqrt{N_s}$ around the original curve, where $\sigma$ is the standard deviation of the $N_s$ samples. The presence of the $\sqrt{N_s}$ in the denominator is due to the fact that the random variable under consideration ($E_\phi$) is itself an average over $N_s$ random variables corresponding to each sample.

\begin{figure}
\begin{center}
\includegraphics[width=0.75\textwidth]{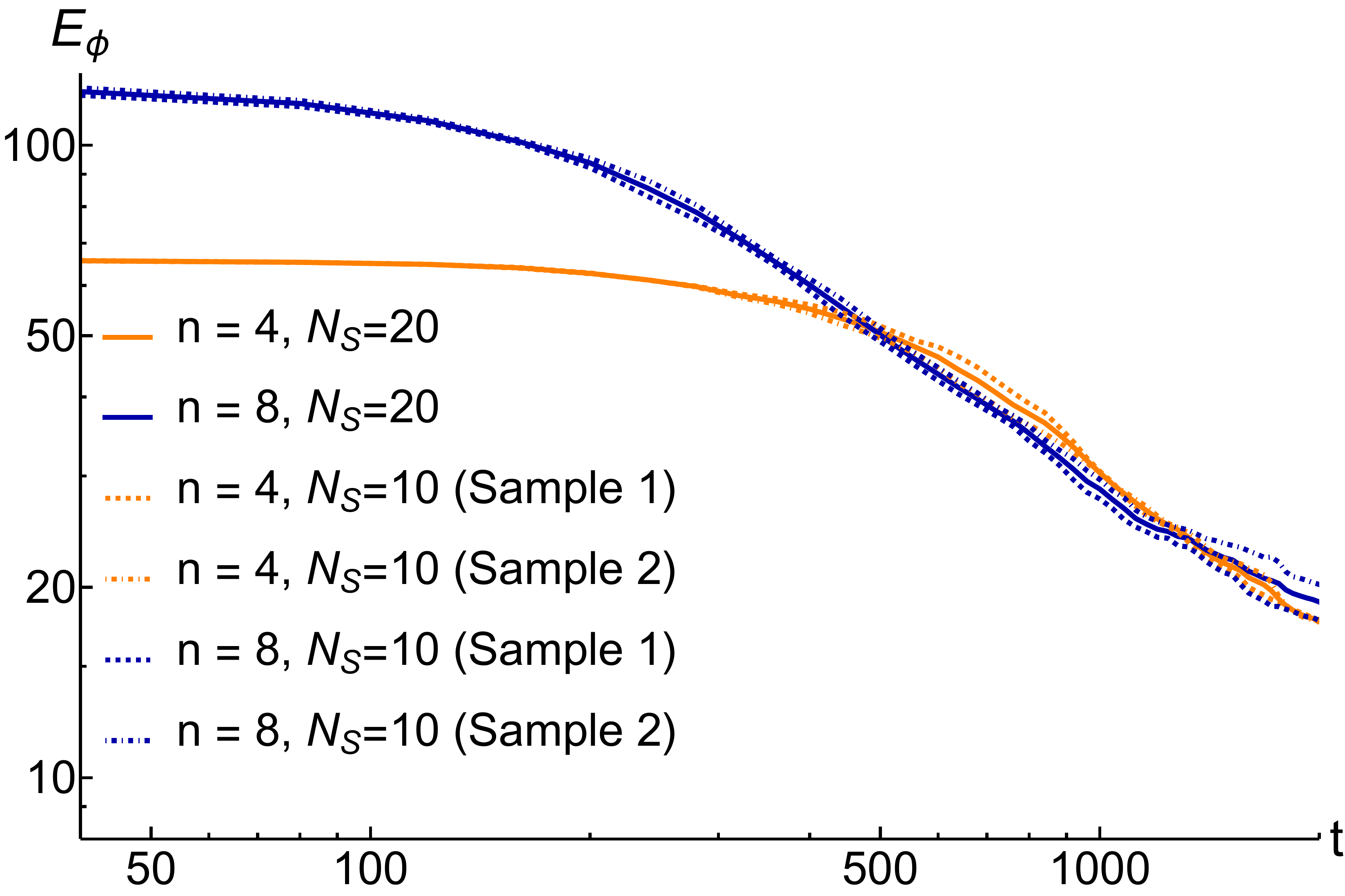}
\caption{\label{fig:comp_n_NS} Time evolution of the energy in the background $E_\phi$ for $n = 4$ (orange) and $n = 8$ (dark blue) for three different samples of kink-antikink ensembles each: two independent samples of $N_S=10$ realizations, and one sample with $N_S=20$ realizations. Here $\lambda=0.3$, $\mu=0.1$, and the units are such that $m=1$.
}
\end{center}
\end{figure}

However we should keep in mind that the above result was obtained for a particular choice of initial number of kink-antikink pairs $n$ as well as for one particular sample of different realizations of the ensemble evolution. In Fig.~\ref{fig:comp_n_NS} we check that the above power law behavior is independent of the particular sample and its size (up to statistical fluctuations), as well as of the initial number of kink-antikink pairs in the ensemble. The fact that the late time decay dynamics does not depend on the initial pair density was to be expected on physical grounds since, if one starts with a higher number of kinks and antikinks, the initial annihilation rate increases, eventually erasing any memory of the starting configuration. The net effect is simply to advance the onset of the power law phase of the decay without changing the exponent.

\begin{figure}
\begin{center}
\includegraphics[width=0.75\textwidth,angle=0]{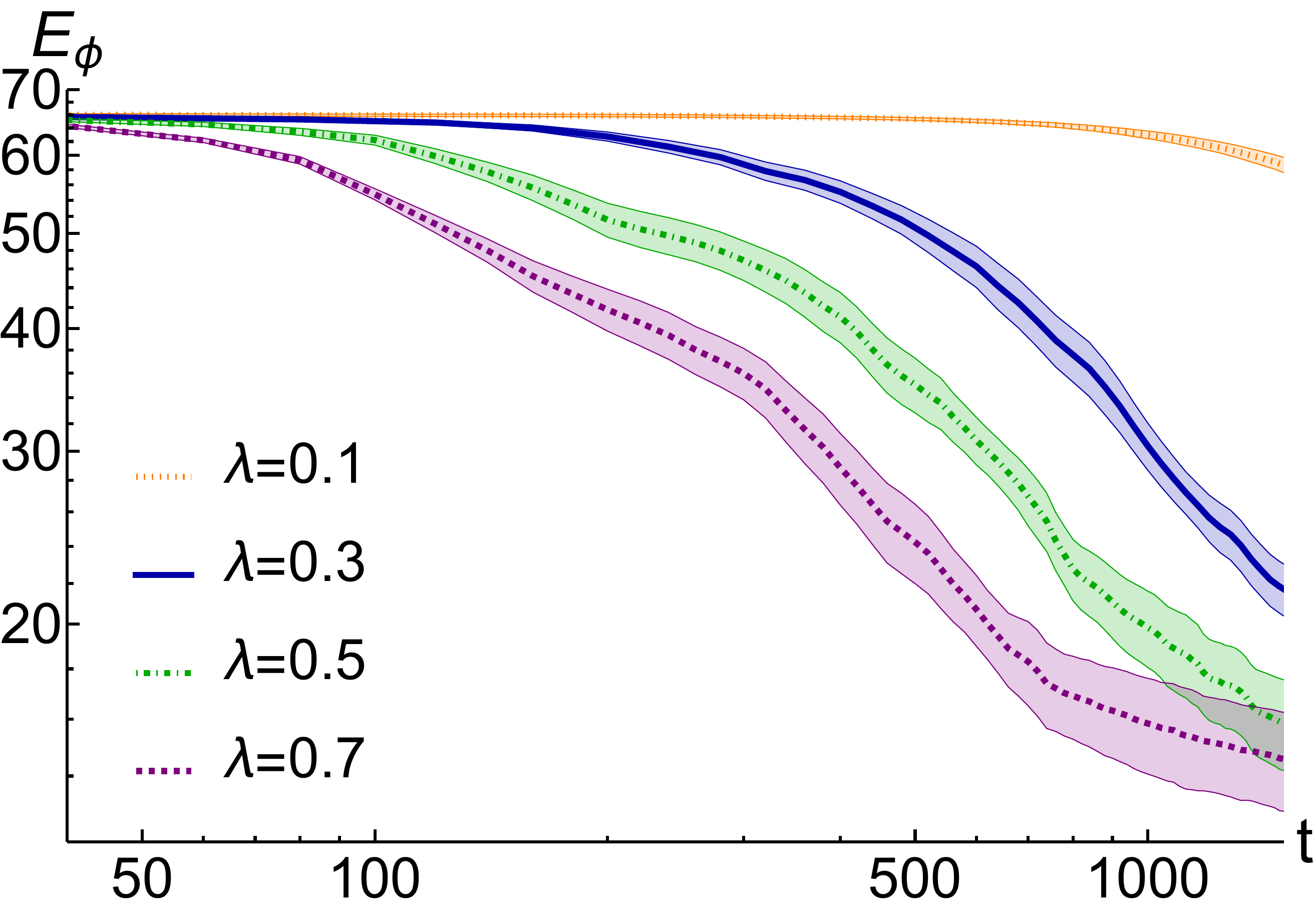}
\caption{\label{fig:lambda_dep} Time-evolution of the energy in the background $E_\phi$ for an ensemble of $n = 4$ kink-antikink pairs averaged over a sample of $N_S=20$ different realizations, for different values of the coupling $\lambda$. Here $\mu = 0.1$ and units are such that $m=1$. The shaded bands show the statistical uncertainty given by $\pm (\sigma/\sqrt{N_s})$, where $\sigma$ is the standard deviation of the $N_s$ samples.
}
\end{center}
\end{figure}

\begin{figure}
\begin{center}
\includegraphics[width=0.75\textwidth,angle=0]{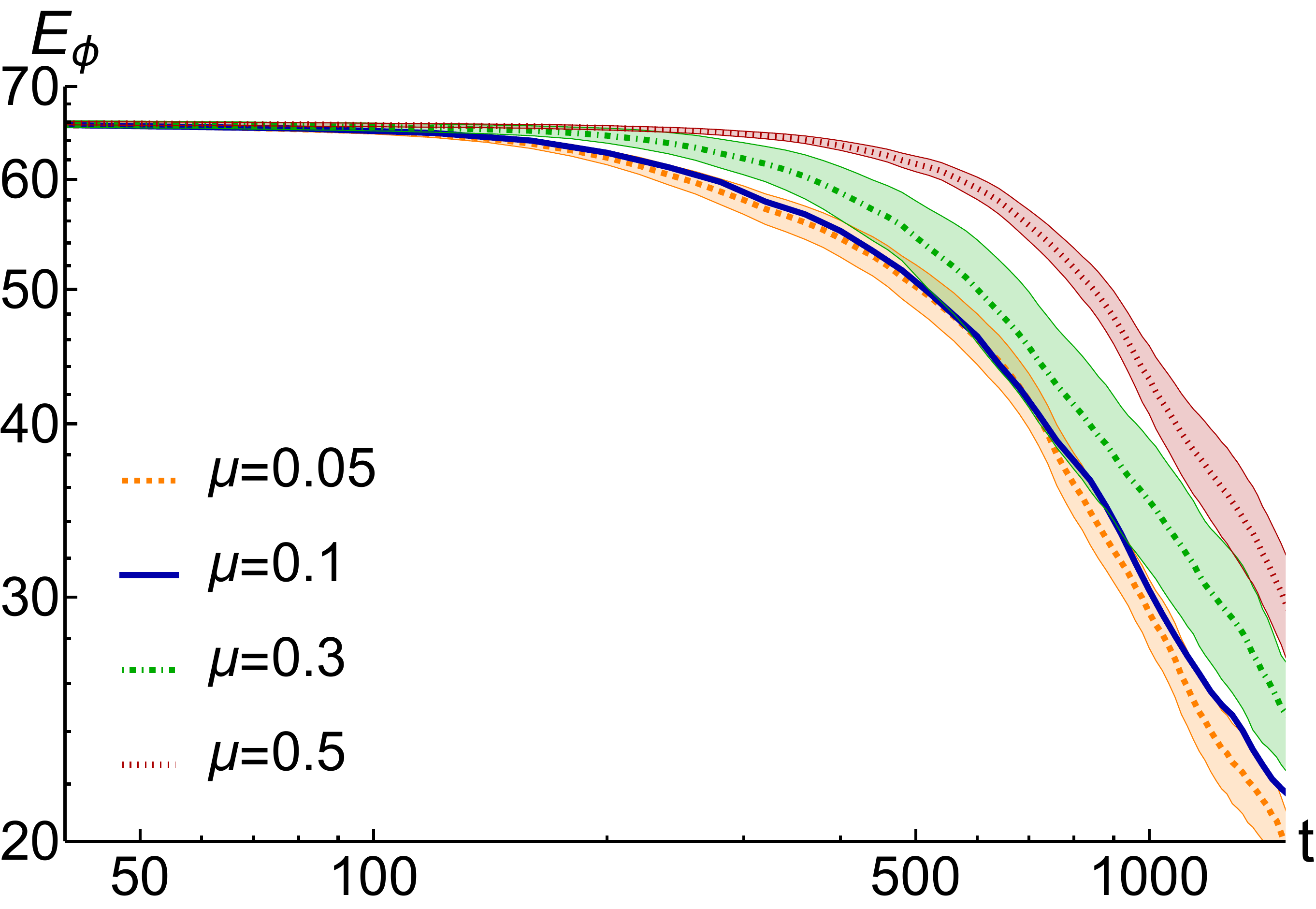}
\caption{\label{fig:mu_dep} Time-evolution of the energy in the background $E_\phi$ for an ensemble of $n = 4$ kink-antikink pairs averaged over a sample of $N_S=20$ different realizations, for different values of the quantum field mass $\mu$. Here $\lambda = 0.3$ and units are such that $m=1$. The shaded bands show the statistical uncertainty given by $\pm (\sigma/\sqrt{N_s})$, where $\sigma$ is the standard deviation of the $N_s$ samples. Note that the statistical uncertainty band for $\mu = 0.1$ (solid dark blue) is the same as in Fig.~\ref{fig:lambda_dep}, so we do not show it here.
}
\end{center}
\end{figure}

It is also interesting to investigate the dependence of the exponent $\alpha$ on the {\it physical} parameters of the model. To do this we consider samples of size $N_S=20$ with $n=4$ initial kink-antikink pairs, and vary the parameters $\lambda$ and $\mu$. As we show in Figs.~\ref{fig:lambda_dep} and~\ref{fig:mu_dep}, to the level of accuracy permitted by our simulations, the late time decay of the kink-antikink ensemble is consistent with a {\it universal} value of the exponent $\alpha(\sim0.6-0.9)$ that is strictly smaller than 1. Indeed, similarly to the analysis of Fig.~\ref{fig:main_res} we can fit a power law to the asymptotic tails of the plots.  We sum up our results in Table~\ref{tab:fits}, where we gather the different exponents $\alpha$ and their uncertainties $\delta\alpha$ for different values of $\lambda$ and $\mu$.

Unfortunately our results are not sufficiently accurate to draw stronger conclusions. In particular, we notice from Fig.~\ref{fig:lambda_dep}, for small values of coupling, $\lambda \lesssim 0.1$, the energy lost by the background is negligible for the duration of our runs ($t = 2000 m^{-1}$). Also, and perhaps more importantly, for higher values of coupling $\lambda \gtrsim 0.7$, the decay is rapid and hence $|t_{\rm start} - t_{\rm end}|$ is small, leading to large relative errors.

\begin{table}[]
    \centering
    \begin{tabular}{|c|c|c|c|c|}
    \hline
$\lambda$     & $t_{\rm start}$ & $t_{\rm end}$ & $\alpha$ & $\delta \alpha$ \\
\hline
    0.1 & - & - & -  & - \\
    0.3 & 560 & 920 & 0.73 & $ 0.22$\\
   0.5  & 340 & 540 & 0.65 & $ 0.24$\\
    0.7 & 320 & 380 & 0.80 & $ 0.71$\\
    \hline
    \end{tabular}
    \hspace{2cm}
    \begin{tabular}{|c|c|c|c|c|}
    \hline
   $\mu$  & $t_{\rm start}$ & $t_{\rm end}$ & $\alpha$ & $\delta \alpha$ \\
   \hline
   0.05	  & 760 & 980 &  0.93 & $ 0.42$\\
   0.1  & 560 & 920 & 0.73 & $ 0.22$\\
   0.3  & 620 & 1080 & 0.71 & $ 0.18$\\
   0.5 & 780 & 1380 & 0.85 & $ 0.20$\\
   \hline
     \end{tabular}
    \caption{Late time power law fit parameters corresponding to the different curves shown in Figs.~\ref{fig:lambda_dep} and~\ref{fig:mu_dep}.}
    \label{tab:fits}
\end{table}
\section{Discussion}
\label{sec:disc}

In this work, we studied the dynamics of classical multiple kink-antikink configurations of a field $\phi$ in $1+1$ dimensions coupled to a quantum field $\psi$. The non-adiabaticity resulting from the collision of kink-antikink pairs result in the excitation of the quantum field and thus produce quantum radiation. This in turn backreacts on the kink-antikink background ensemble and leads to its decay. Our main goal was to quantify the rate of quantum decay of the background at late times and investigate whether a scaling regime sets in.

The full quantum problem including the effects of backreaction of the quantum field $\psi$ on the classical background $\phi$ was tackled by discretizing the field theory model and using the semiclassical approximation. This boiled down to solving the coupled system of differential equations~\eqref{eq:discsGbarbis} and~\eqref{eq:CQCeqs} with initial conditions given by~\eqref{eq:phiIC},~\eqref{eq:phidotIC} and~\eqref{eq:CQCics}. A representative example of an initial configuration of the background kink-antikink ensemble $\phi_0(x)$ along with the initial (renormalized) energy density in $\psi$ are shown in Figs.~\ref{fig:phibkg} and~\ref{fig:rho}. In Fig.~\ref{fig:energy_1sim} we show the time evolution of the energies in the background ($E_\phi$) and the quantum field ($E_\psi$) starting from this initial configuration. Of course the energy lost by the background is equal to the energy gained by the quantum field. 

As can be seen in the figure, the decay does not occur in a smooth way since there are only 4 kink-antikink pairs and every time a pair annihilates there is a noticeable jump in the energy. In order to be able to identify a scaling regime for the late time quantum decay of the kink-antikink ensemble, one would need to consider configurations with a large number of pairs and be able to evolve them for a long enough time. Given the computational complexity, this is problematic and we therefore opt for an alternative route. We evolve samples of many ($N_S=20$) small-number-of-pairs ($n=4$ or $n=8$) configurations (with random initial relative positions and velocities) and average over the background energies to obtain a smoothed out behavior of $E_\phi$. Our main result (for $n=8$ pairs) is shown in Fig.~\ref{fig:main_res}, where we see that at late times ($t\gtrsim 300 m^{-1}$), $E_\phi \propto t^{-\alpha}$ with $\alpha$ strictly smaller than 1. The error on the exponent $\alpha$ is estimated from the variance in the values of $E_\phi$ between different realizations (see Eq.~\ref{eq:error}). We also check that the above late time behavior doesn't depend on the number of pairs or the particular choice of sample of realizations (see Fig.~\ref{fig:comp_n_NS}). 

It is important to mention here that we cannot strictly trust the simulations after the background has lost more than half its initial energy because by that time the assumption that the background is classical presumably ceases to be true, and the semi-classical approximation breaks down. In other words, in that regime the separation between classical and quantum degrees of freedom becomes ambiguous. At the same time, we are interested in identifying the \emph{late time} scaling of $E_\phi$. Together, these create a tension and constitute a serious limitation to our current work, leading to large sources of uncertainty on the scaling exponent $\alpha$.

The above limitation becomes particularly clear when investigating how the scaling exponent $\alpha$ depends on the coupling or the mass of the quantum field. The time evolution of $E_\phi$ (with $n=4$ pairs) for different values of $\lambda$ and $\mu$ is shown in Figs.~\ref{fig:lambda_dep} and~\ref{fig:mu_dep}, while the corresponding values of $\alpha$ and $\delta\alpha$ are assembled in Table~\ref{tab:fits}. We find that the value of $\alpha$ is consistent with a value strictly smaller than 1 and independent of the physical parameters. However, we note that, because of our numerical and statistical limitations, we cannot yet conclude as to the universality of the scaling exponent. 

Given the $\mathcal{O}(N^2)$ complexity of the problem, we are limited by both resolution and statistics. We attempt to navigate the former by ensuring our results are independent of the lattice parameters $N$ and $L$ at least for the relevant time durations (see Fig.~\ref{appfig:ndep}), while we mitigate the latter by using a sample size of $N_s = 20$. The current results are exploratory in nature and future work will involve achieving better resolutions and using larger samples with more kink-antikink pairs ($n = 16, 32, \dots$). Note that since the initial kink and antikinks need to be non-interacting this would require a larger lattice size $L$. To maintain sufficient resolution a larger $L$ would also entail a larger $N$, significantly adding to the computational cost. A lattice of size $L \sim 2000$ with sufficiently high resolution ($N\sim 5000$) containing $n \sim 100$, well-separated kink-antikink pairs would be ideal to verify the preliminary results presented in this work.
Of course a much more interesting setting would be to study the quantum annihilation of domain wall networks in higher dimensions ($2+1$ or $3+1$), but this would be even more computationally intensive and would require parallelization. 

One of the interesting physical applications of the current work is the production of dark matter, in particular, super-heavy dark matter (SHDM) in the early universe as a result of the collision of domain wall networks~\cite{Kitajima:2022lre,Lee:2024xjb} (or bubble walls~\cite{Falkowski:2012fb}). SHDM involves dark matter particles with masses $10^4\ {\rm GeV} \lesssim m_\chi \lesssim 10^{19}\ {\rm GeV}$, which are out of reach in current particle colliders. But in some theories of physics beyond the Standard Model, the electroweak phase transition in the early universe could produce bubble walls which expand rapidly and collide. These bubble walls in the absence of hydrodynamic obstacles in the plasma can become ultra-relativsitic~\cite{Espinosa:2010hh} before colliding. With bubble walls colliding, the energy stored in the walls is transferred to the surrounding plasma. For typical runaway bubble walls (Lorentz factor $\gamma \lesssim 10^{15}$) this energy can be large and in the presence of a sufficiently large coupling with dark matter can produce SHDM with the correct relic abundance. Analogously to the current work, the classical kink-antikink ensemble background could be extended to an ultra-relativistic domain wall or bubble wall network, and the quantum $\psi$ particles would correspond to the SHDM candidate. Although further investigations are necessary, the toy model discussed here could thus be interpreted as a proof of concept for the full physically relevant simulation where generically classical decay due to self-annihilation will compete with the quantum decay mode studied in this work.
\acknowledgments
We thank Tanmay Vachaspati for comments and careful reading of our manuscript. 
We wish to thank the Yukawa Institute for Theoretical Physics, Kyoto University, the Tokyo Electron House of Creativity, Tohoku University, for hospitality during the period when this project was initiated. 
M.\,M. and K.\,M. are supported by NSF Grant No. AST-2108466. M.\,M. also acknowledges support from the Institute for Gravitation and the Cosmos (IGC) Postdoctoral Fellowship. 
The work of K.M. is supported by the NSF Grant Nos.~AST-2108467, and AST-2308021, and KAKENHI No.~20H05852. 
The work of A.N. was partly supported by JSPS KAKENHI Grant Numbers 20H05852, JP19H01891, JP23H01171, 23K25868. A.N. thanks to the molecule workshops ``Revisiting cosmological non-linearities in the era of precision surveys” YITP-T-23-03 and ``Extreme Mass Dark Matter Workshop: from Superlight to Superheavy” YITP-T-23-04 since discussions during the workshop were useful for this work. 
G.\,Z. is supported by SONATA BIS grant 2023/50/E/ST2/00231 from the Polish National Science Centre. The data used to produce the figures in this work are available at~\cite{UJ/9QPB8C_2025}.
\appendix
\section{Spatial resolution}
\label{appsec:numerical_tech}
\begin{figure}
\begin{center}
\includegraphics[width=0.73\textwidth,angle=0]{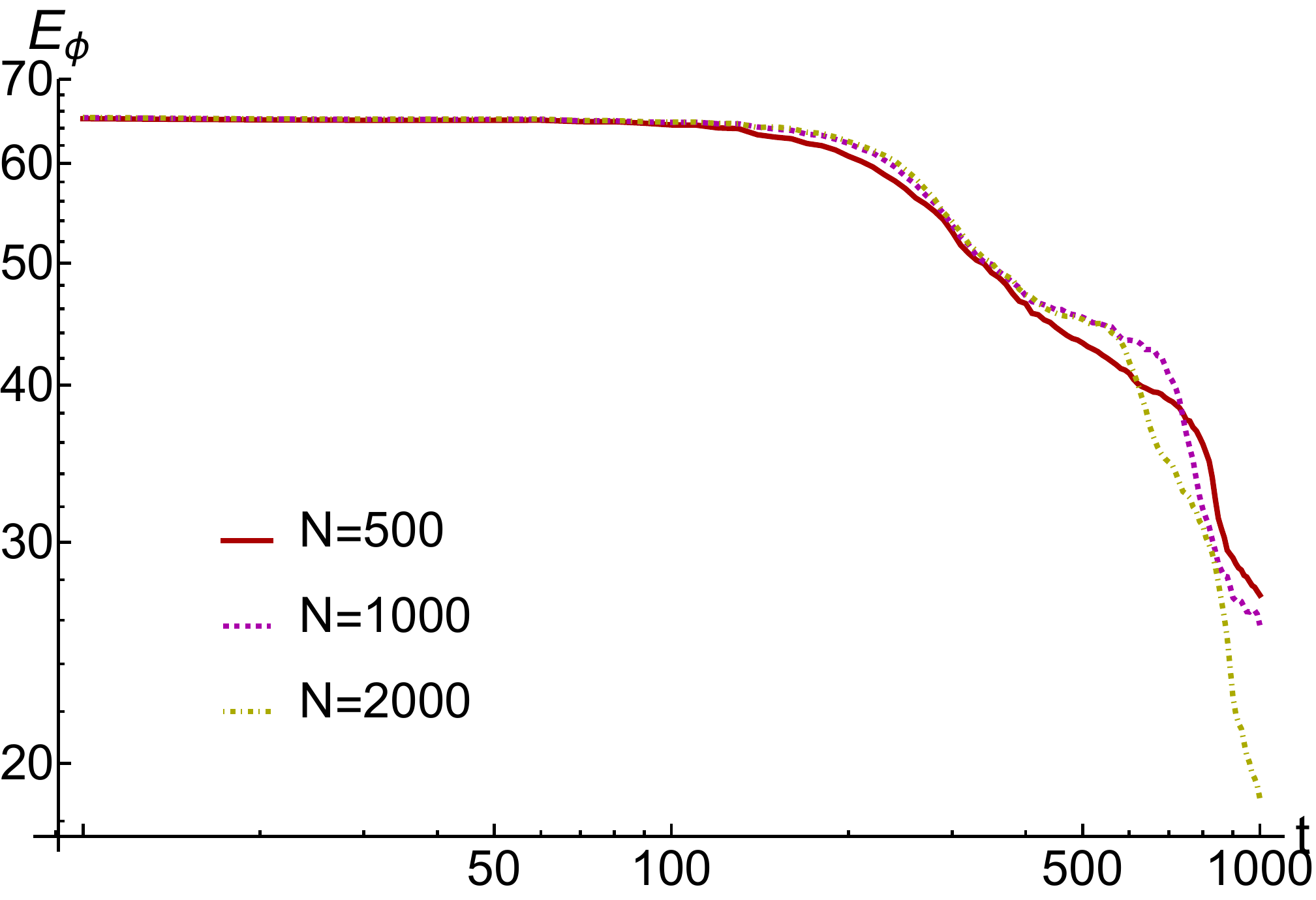}
\caption{\label{appfig:ndep} Time-evolution of the energy in the background $E_\phi$ for $N = 500$ (solid dark orange), $1000$ (dashed dark magenta), and $2000$ (dot-dashed dark yellow) for one realization. Here $\lambda=0.3$, $\mu=0.1$, and the units are such that $m=1$.
}
\end{center}
\end{figure}

In this appendix we discuss the sensitivity of our results to the spatial resolution. While we need to choose a small enough lattice spacing $a=L/N$ to resolve individual kinks and antikinks, at fixed $L$ this requires a large enough number of lattice points $N$. This number is however limited by the $\mathcal{O}(N^2)$ computational complexity. We therefore need to make sure that our fiducial choice of $N=500$ is accurate enough. In Fig.~\ref{appfig:ndep}, we plot the energy in the background $E_\phi$, one of the main observables we are interested in, for different values of $N$ (500,1000, and 2000) keeping $L=200$ fixed (and with numerical integration time-step\footnote{We have also checked that the total energy is conserved to within $1\%$ for the chosen fiducial value of $dt = a/5$.} $dt = a/5$). We note that the three curves start deviating for $t \gtrsim 500 m^{-1}$, but the \emph{relative error} stays within $\sim 10\%$ by the time $E_\phi$ reaches half its initial value and the semi-classical approximation ceases to be valid. While this clearly shows $N=500$ is not the ideal choice, it offers a fair trade-off between numerical accuracy and computational complexity. 
\bibliography{refs}
\bibliographystyle{jhep}
\end{document}